\documentclass[10pt,twocolumn,french,english]{revtex4}
\usepackage[T1]{fontenc}
\usepackage[latin9]{inputenc}
\usepackage[a4paper]{geometry}
\usepackage{array}
\usepackage{amsmath}
\usepackage{graphicx}
\usepackage{amssymb}
\usepackage{esint}

\makeatletter

\providecommand{\tabularnewline}{\\}

\@ifundefined{textcolor}{}
{%
 \definecolor{BLACK}{gray}{0}
 \definecolor{WHITE}{gray}{1}
 \definecolor{RED}{rgb}{1,0,0}
 \definecolor{GREEN}{rgb}{0,1,0}
 \definecolor{BLUE}{rgb}{0,0,1}
 \definecolor{CYAN}{cmyk}{1,0,0,0}
 \definecolor{MAGENTA}{cmyk}{0,1,0,0}
 \definecolor{YELLOW}{cmyk}{0,0,1,0}
 }

\makeatother

\usepackage{babel}
\addto\extrasfrench{\providecommand{\fg}{\ifdim\lastskip>\z@\unskip\fi~\frqq}}

\begin{document}

\title{Experimental demonstration of Aharonov-Casher interference in a \\
Josephson junction circuit}

\author{I. M. Pop$^{1}$, B. Douçot$^{2}$, L. Ioffe$^{3}$, I. Protopopov$^{4}$,
F. Lecocq$^{1}$, I. Matei$^{1}$, O. Buisson$^{1}$ and W. Guichard$^{1}$\smallskip{}
}

\affiliation{{\small $^{1}$ Institut Néel, CNRS et Université Joseph Fourier,
BP 166, F-38042 Grenoble Cedex 9, France}}

\affiliation{{\small $^{2}$ Laboratoire de Physique Théorique et Hautes Energies,
CNRS UMR 7589, Universités Paris 6 et 7, 4 place Jussieu, 75005 Paris,
France}}

\affiliation{{\small $^{3}$Department of Physics and Astronomy, Rutgers University,
136 Frelinghuysen Rd., Piscataway, NJ 08854, USA}}

\affiliation{{\small $^{4}$L. D. Landau Institute for Theoretical Physics, Kosygin
str. 2, Moscow 119334, Russia and Institut fuer Nanotechnologie, Karlsruher
Institut fuer Technologie, 76021 Karlsruhe, Germany\smallskip{}
}}
\begin{abstract}
A neutral quantum particle with magnetic moment encircling a static
electric charge acquires a quantum mechanical phase (Aharonov-Casher
effect). In superconducting electronics the neutral particle becomes
a fluxon that moves around superconducting islands connected by Josephson
junctions. The full understanding of this effect in systems of many
junctions is crucial for the design of novel quantum circuits. Here
we present measurements and quantitative analysis of fluxon interference
patterns in a six Josephson junction chain. In this multi-junction
circuit the fluxon can encircle any combination of charges on five
superconducting islands, resulting in a complex pattern. We compare
the experimental results with predictions of a simplified model that
treats fluxons as independent excitations and with the results of
the full diagonalization of the quantum problem. Our results demonstrate
the accuracy of the fluxon interference description and the quantum
coherence of these arrays.
\end{abstract}
\maketitle
The formation of macroscopically large coherent states in systems
with many unquenched degrees of freedom tests our understanding of
quantum mechanics and it is essential for quantum computation. One
of the most striking consequences of such coherence are interference
patterns that are expected to appear when a charged particle encircles
a magnetic flux (Aharonov-Bohm effect\cite{Aharonov1959}) or when
a flux encircles a charge (Aharonov-Casher effect\cite{Aharonov1984}).
The quantum coherence implied by these effects is a fragile phenomenon
which is easily destroyed by uncontrolled degrees of freedom. In an
ideal Josephson junction array most microscopic degrees of freedom
are quenched by electron pairing into Cooper pairs; the only remaining
degrees of freedom are the phase of the order parameter on each island
or the charge conjugated to it. Observation of the interference provides
the evidence of the full control of the quantum system in the ground
state. In Josephson junction arrays it proves the irrelevance of the
uncontrolled degrees of freedom such as two level systems, non-equilibrium
quasi-particles, etc.

Aharonov-Bohm (AB) effect in small and large Josephson arrays is a
very well established phenomenon: in the former it leads to critical
current oscillations in SQUIDs\cite{ClarkeBraginskiBook}, in the
latter it results in a complicated magnetic field dependence with
many peaks at commensurate fields\cite{Voss1982,Voss1983,Chen1996,Fazio2001}.
The experimental confirmation of its dual, the Aharonov-Casher (AC)
effect, is less clear. It was observed for small Josephson circuits
where vortices moved in a ring encircling a single charge\cite{Elion1993}.
However, large arrays studied in a number of works show the appearance
of the intermediate {}``normal'' phase of the arrays which is characterized
by a non-zero resistance\cite{VanderZant1993,Chen1996,SerretThesis2002}.
Non-zero resistance implies that the fluxon motion is dissipative;
this excludes quantum coherence. It is very important to establish
the presence or absence of this dissipation and its possible origin
in well controlled medium size arrays. This is the main goal of our
work.

The duality of AB and AC effects can be illustrated by analyzing the
quantum mechanical phase resulting from the braiding of particle with
charge $q$ and a neutral particle with magnetic moment $\vec{\mu}$.
If the charged particle is at rest while the neutral one moves, the
former generates an electrical field $\vec{E_{\mu}}$ at the position
$\vec{R_{\mu}}$ of the latter that gives the interaction energy $I_{1}=[\vec{\frac{\mu}{c}}\times\vec{E_{\mu}}]\cdot\dot{\vec{R_{\mu}}}$.
Conversely, the magnetic moment at rest generates a vector potential
$\vec{A_{q}}$ at the position $\vec{R_{q}}$ of the moving charge,
that gives the interaction energy $I_{2}=\frac{q}{c}\,\vec{A_{q}}\cdot\dot{\vec{R_{q}}}.$
In either case, the acquired phase is given by the time integral of
the interaction energy. In case of the moving charge, this phase is
$\delta\phi_{AB}=\left(\frac{q}{hc}\right)\oint\vec{A_{q}}\cdot\vec{dR_{q}}$
(AB effect); in case of a moving magnetic moment this phase is $\delta\phi_{AC}=\frac{1}{hc}\oint\left(\vec{\mu}\times\vec{E_{\mu}}\right)\cdot\vec{dR_{\mu}}$(AC
effect). Experimentally the former was first observed 50 years ago
as an electron interference pattern in magnetic field \cite{Chambers1960};
the latter was measured with percent accuracy in neutron and atomic
interferometry experiments \cite{Cimmino1989,Sangster1993}. In Lorentz
invariant systems of neutral and charged particles the distinction
between the two effects is impossible. What seems as AB effect for
the observer in the rest frame of the neutral particle becomes AC
effect for the observer in the rest frame of the charged particle.
In solid state devices these effects are distinguishable because the
rest frame is fixed by the device, and therefore the observation of
AB interference does not imply the one of AC and vice versa. Because
uncontrolled degrees of freedom turn out to be mostly electric charges
(either background charges or non-equilibrium quasi-particles in superconductors),
experimentally it is difficult to remain in the rest frame of the
charge and, consequently, the observation of the AC effect is much
more challenging.

A Josephson junction circuit can be described either in terms of the
superconducting phase or in terms of the charges of its islands. If
the charging energy $E_{C}=e^{2}/(2C)$ is larger than the Josephson
energy $E_{J}$, Cooper pairs are almost localized. The dynamics of
a Josephson junction array can be viewed as due to the rare motion
of these pairs. In the opposite limit $E_{J}\gg E_{C}$ the charge
is delocalized and the array dynamics can be viewed as due to rare
phase slips resulting in the motion of fluxons. The nature of the
elementary excitations does not preclude however the description of
the circuit in the charge (or phase) basis. The elementary excitations
only become more complicated objects for $E_{J}>E_{C}$ in the charge
basis or for $E_{C}>E_{J}$ in the phase basis.

In the following, we analyze the ground state properties of a 6 Josephson
junction chain as a function of the gate voltage that induces polarization
charges for $E_{J}/E_{C}=2-3$. Because in this regime the individual
excitations are fluxons, the properties of the circuit are due to
the interference between phase-slips on different junctions. The interference
pattern is due to the charges induced on the array islands which is
exactly AC effect. The difference between the longer chain (of six
junctions) studied here and the previous works\cite{Elion1993,Friedman2002}
is that fluxons can take one of the six possible routes resulting
in a much more complicated interference. In the following we compute
the expected properties of the array assuming that phase slips are
independent excitations. Because this assumption might be questioned
for $E_{J}/E_{C}=2-3$ we have also performed the complete diagonalization
of the Hamiltonian. Finally we compare the results of both approaches
to the measured data. Our main conclusion is that the phase slip approximation
provides a semi-quantitative description of the data in this regime
and that the observed interference pattern evidences the quantum coherent
dynamics of our relatively large circuits.

\begin{figure}[h]
\begin{centering}
\includegraphics[width=0.85\columnwidth]{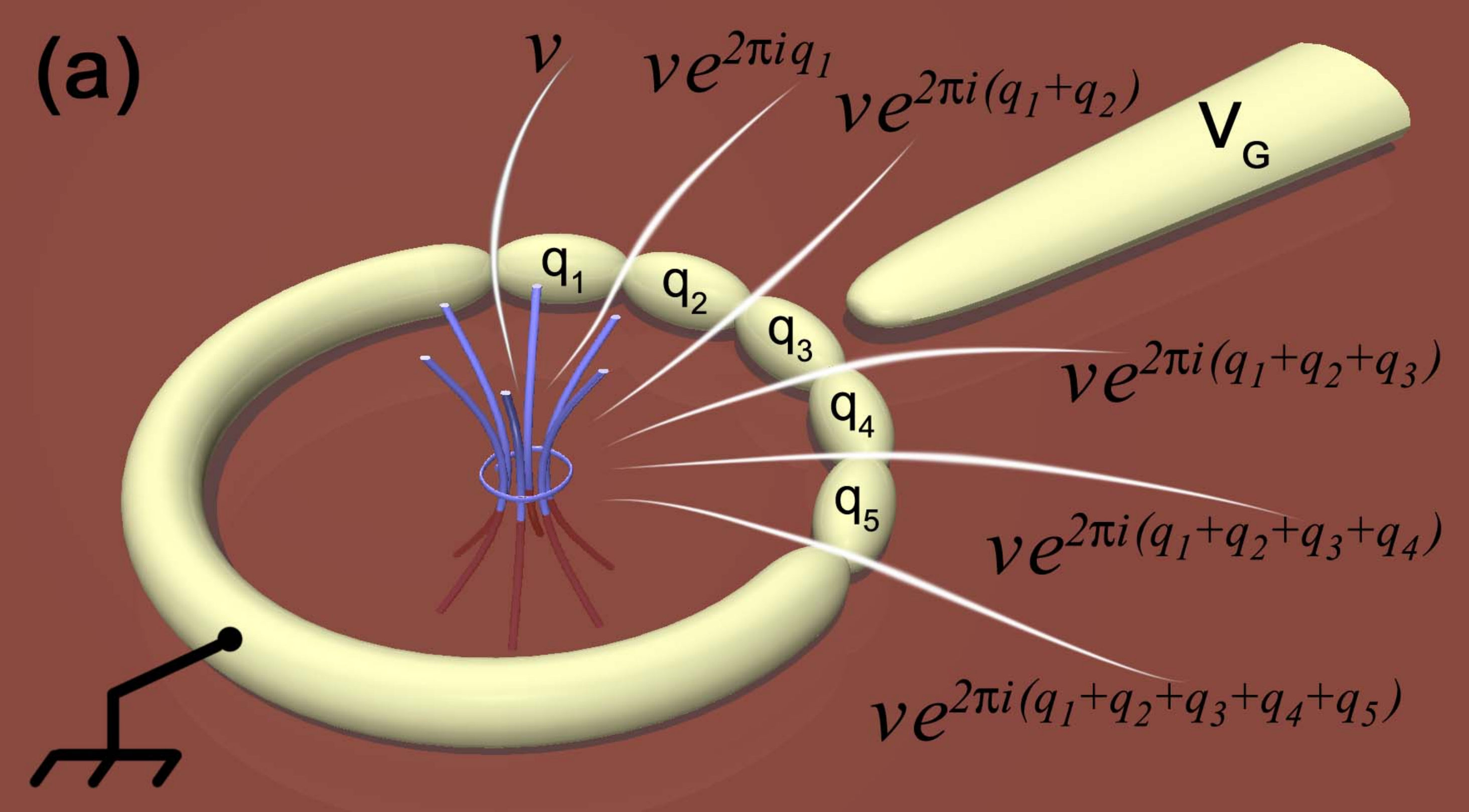}
\par\end{centering}

\smallskip{}

\begin{centering}
\includegraphics[width=0.9\columnwidth]{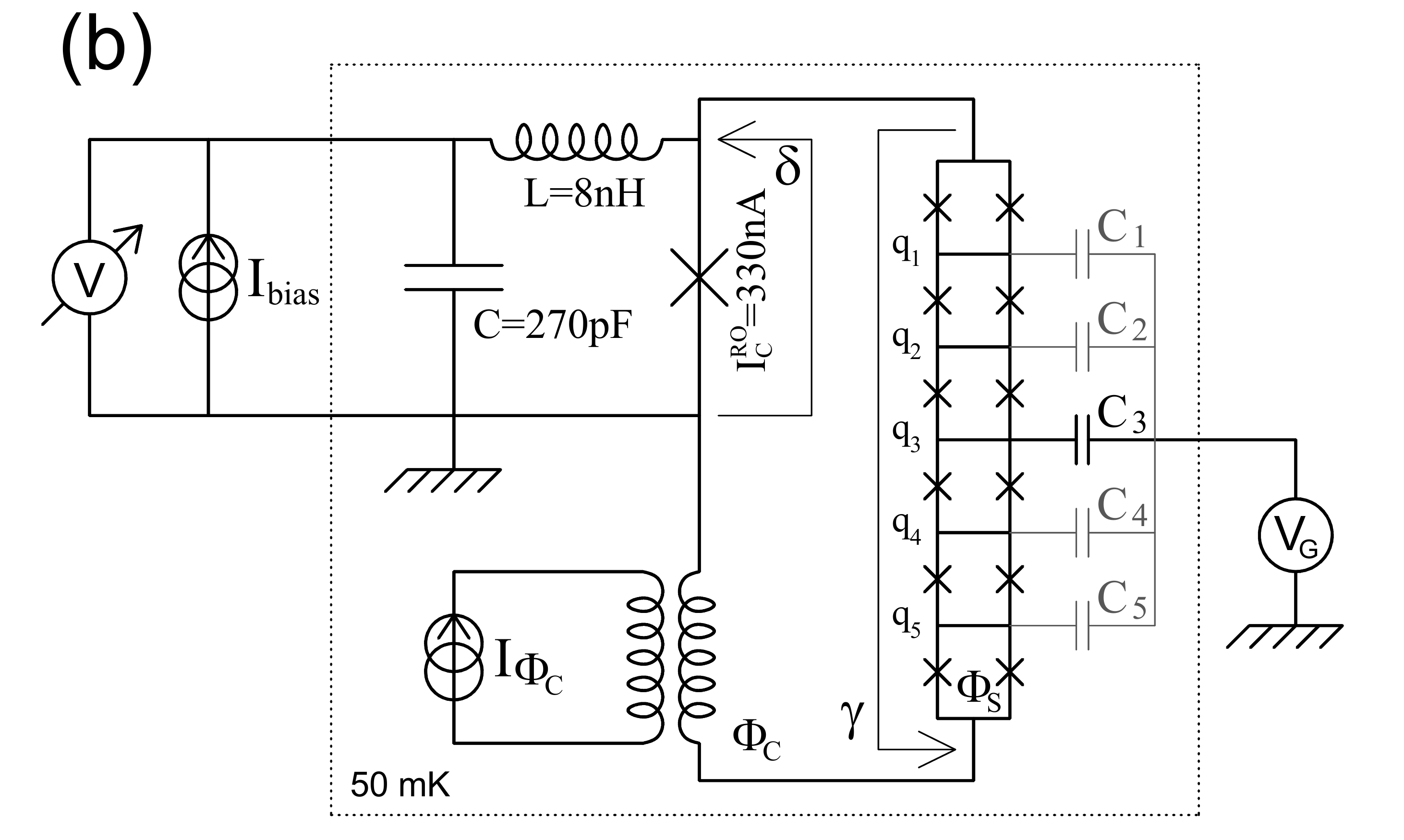}
\par\end{centering}

\begin{centering}
\includegraphics[width=0.9\columnwidth]{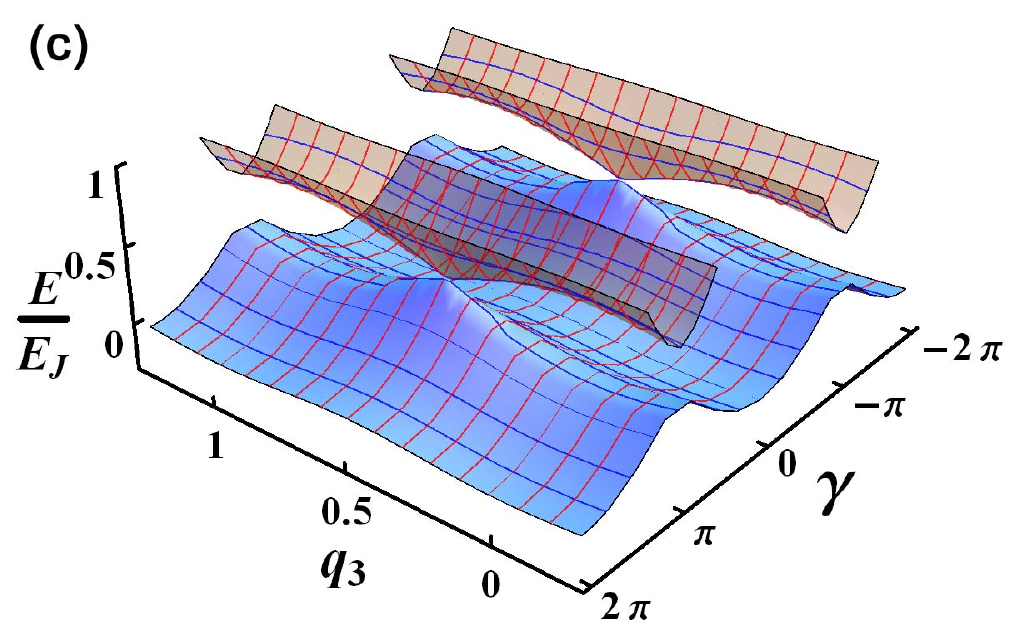}
\par\end{centering}

\caption{Schematic view of the experimental setup used to probe the phase-slip
interference in a chain of 6 Josephson junctions. In (a) we show an
idealized view of the experimental design. The chain contains five
small superconducting islands connected to each other and to the leads
by identical Josephson junctions. The islands are coupled to a nearby
gate electrode. In (b) we present the electrical scheme of the measurement.
The six-SQUID chain is inserted in a superconducting loop. The flux
$\Phi_{C}$ created by on-chip coils controls the phase difference
$\gamma$ over the chain. The independently controlled flux $\Phi_{S}$
through the SQUID loops is used to tune \textit{in situ} the Josephson
coupling $E_{J}=E_{J}^{0}\cos\left(\pi\Phi_{S}/\Phi_{0}\right)$,
where $E_{J}^{0}=2\,\mbox{K}$ and $\Phi_{0}$ is the magnetic flux
quantum. The charging of one SQUID is $E_{C}=660\,\mbox{mK}.$ The
phase difference over the read-out junction is denoted by $\delta$.
The gate electrode couples to the charge $q_{i}$ on island $i$ via
the capacitance $C_{i}^{g}$. The coupling to the central island $C_{3}^{g}$
is at least 10 times larger than all other capacitances and determines
the dominant gate effect at low voltage. (c) shows the calculated
ground and first excited state for the 6 junction chain as a function
of the phase bias $\gamma$ and the induced charge $q_{3}$ on the
central island, for charge configurations of the type $\left(0,0,q_{3},0,0\right)$.}

\label{RingPlusVortex}
\end{figure}

Fig.\ref{RingPlusVortex}a shows an idealized view of our circuit:
a superconducting ring containing five islands connected by Josephson
junctions. A gate voltage $V_{G}$ induces the charge frustration
$q_{i}=C_{i}^{g}V_{G}/(2e)+q_{i}^{0}$ on the \textit{i}-th island.
Here $q_{i}^{0}$ are static offset charges. The couplings to the
gate electrode $C_{i}^{g}$ are not equal for all islands and induce
a general charge configuration $\kappa=(q_{1},q_{2},q_{3},q_{4},q_{5})$,
expressed in units of $2e$. White traces in Fig.\ref{RingPlusVortex}a
represent six possible paths for a fluxon to cross the ring, through
one of the six Josephson junctions. The probability of this event
is given by the quantum phase-slip amplitude $\upsilon_{j}$ of a
single junction. $\upsilon_{j}$ contains an AC phase-factor depending
on the islands charges $q_{i}$ (see eq. (\ref{eq:QPS}) and (\ref{eq:MGLwidthOfBlochBand})).
The ground state of the SQUID chain depends on the Coherent Quantum
Phase-Slip (CQPS) amplitude that results from the macroscopic interference
of six fluxons. The CQPS amplitude, $v^{*}$, is obtained by summing
up all phase-slip amplitudes on the junctions \cite{Matveev2002}
(for computation details see the Supplementary Information text):

\begin{equation}
v^{*}=\sum_{j=1}^{6}v_{j}\quad\mbox{and}\quad v_{j}=v\, exp\left[i2\pi\sum_{k=1}^{j-1}q_{k}\right]\label{eq:QPS}\end{equation}

Here $\upsilon$ is the magnitude of the phase-slip amplitude for
a single Josephson junction. In the quasi-classical approximation,
valid at $E_{J}\gg E_{c}$, it is\cite{Likharev1985}:

\begin{equation}
v=8\sqrt{\frac{E_{J}E_{C}}{\pi}}\left(\frac{E_{J}}{2E_{C}}\right)^{0.25}e^{-\sqrt{8\frac{E_{J}}{E_{C}}}}\label{eq:MGLwidthOfBlochBand}\end{equation}

The first two energy levels of a Josephson junction chain are shown
in Fig. \ref{RingPlusVortex}c. These energy levels have been calculated
by diagonalizing the Matveev-Larkin-Glazman (MLG) tight-binding Hamiltonian
\cite{Matveev2002}:

\begin{equation}
H\,|m\rangle=E_{m}\,|m\rangle-v^{*}\left[|m-1\rangle+|m+1\rangle\right]\label{eq:TightBindingHamiltonian}\end{equation}
Here $E_{m}=\frac{E_{J}}{2N}(\gamma-2\pi m)^{2}$ is the energy of
the $|m\rangle$ state of the chain polarized at phase $\gamma$ and
$m$ is the quantum variable that counts the number of vortices having
crossed the chain through one of the junctions. The model (\ref{eq:TightBindingHamiltonian})
makes two important assumptions: the quantum phase slips on different
junctions lead to the same quantum states and these tunneling processes
are independent events. As it can be seen from eq. (\ref{eq:QPS}),
the AC interference of CQPS is an intrinsically $2e$ periodic effect.

In our sample, each junction of the chain is realized by a SQUID (see
Fig. \ref{RingPlusVortex}b) to enable tunable Josephson coupling
$E_{J}$. Consequently we can control \textit{in-situ} the strength
of the quantum phase slip amplitude $v$ through the magnetic flux
$\Phi_{S}$. To measure the CQPS effect on the ground state of a Josephson
junction chain, we have shunted the chain by a large read-out junction
(see Fig. \ref{RingPlusVortex}b, \cite{Vion2002} and \cite{Pop2010}).
The flux $\Phi_{C}$ in the superconducting loop containing the read-out
junction and the chain, enables the control of the bias phase $\gamma=\Phi_{C}-\delta$
over the chain. $\delta$ is the phase difference on the read-out
junction.

\begin{figure}[tbph]
\begin{centering}
\includegraphics[width=0.9\columnwidth]{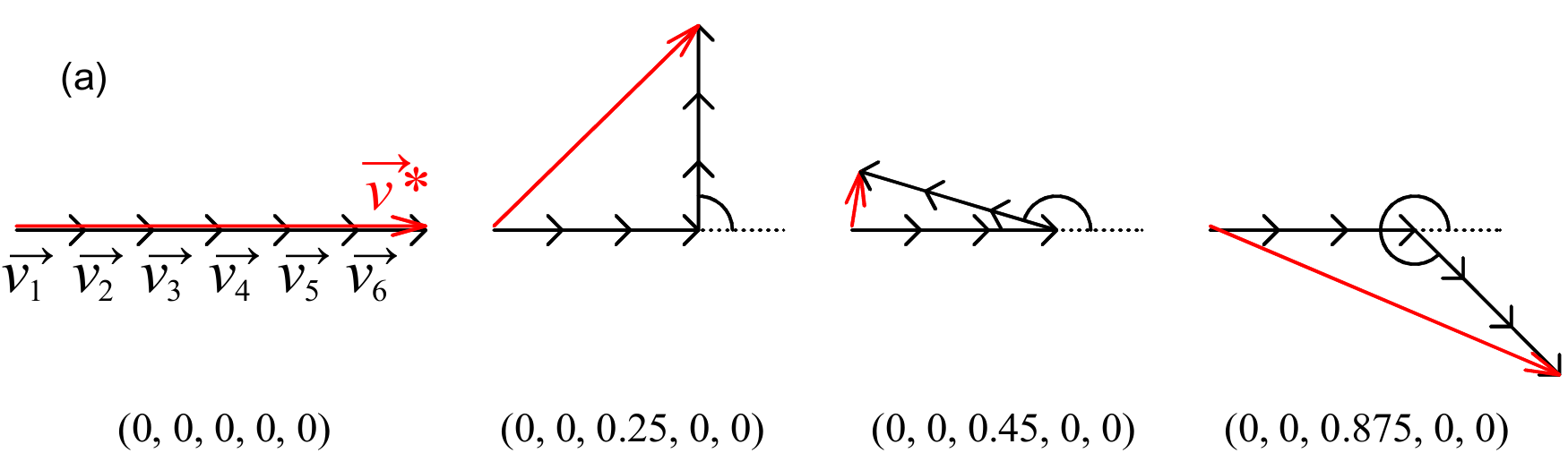}
\par\end{centering}

\begin{centering}
\includegraphics[width=0.9\columnwidth]{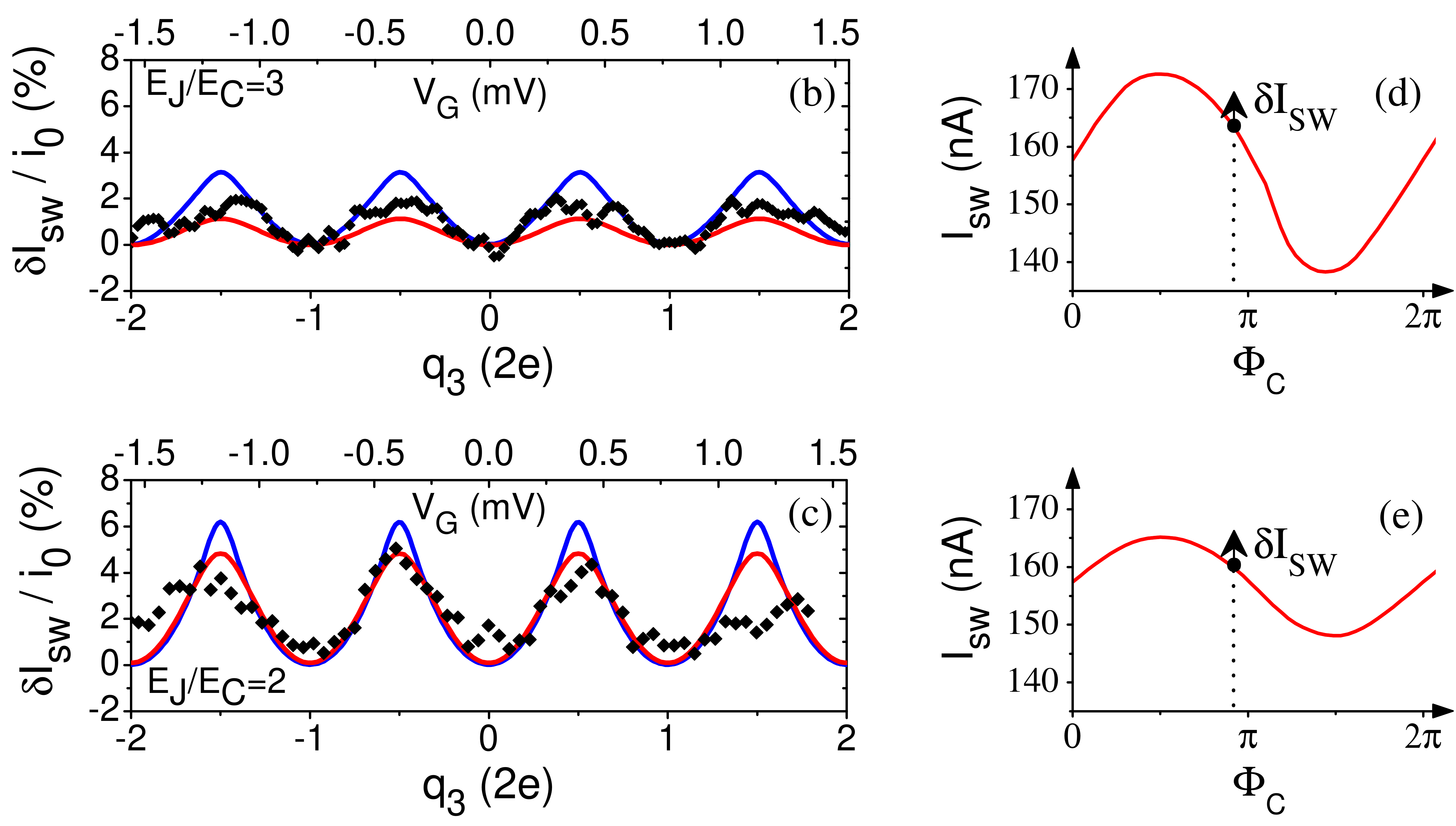}
\par\end{centering}

\caption{Phase-slip interferences controlled by the polarization charge on
the middle island of a 6 Josephson junction chain, corresponding to
the charge configurations $\kappa'=(0\pm0.1,0\pm0.1,\frac{C_{3}^{g}V_{g}}{2e},0\pm0.1,0\pm0.1)$.
(a) Schematic representations of the total CQPS amplitude $v^{*}$
(in red), obtained by summing up the 6 phase-slip probability amplitudes
$v_{i}$ (in black), represented as vectors in the complex plane.
$v^{*}$ is presented for four different charge configurations. (b)
(c) The black diamonds represent the measured variation of the switching
current as a function of the induced charge on the middle island,
in the case of $E_{J}/E_{C}=3$ (b) and $E_{J}/E_{C}=2$ (c). The
\textit{y-axis} is reported in units of the critical current of one
junction, $i_{0}$. The red lines represent the theoretical calculations
in the independent phase slip approximation (MLG model) of CQPS interference.
The chain was phase biased at a constant phase: $\Phi_{C}=0.9\,\pi$.
The working points for the measurements presented in (b) and (c) are
presented in the panels (d) and (e) at the right of each curve. All
curves are shifted so that zero of the $y$ axis corresponds to the
switching current of the zero charge configuration $(0,0,0,0,0)$.
The blue lines represent the corresponding calculations using the
diagonalization of the Hamiltonian (\ref{eq:TotalHamiltonianJosephsonChain}).
There is a reasonable agreement between the semi-classical MLG model
of CQPS interference, numerical calculation and data.}

\label{CentralIslandOscilFig}
\end{figure}

We have measured the switching current (see the Methods section) of
the entire Josephson junction circuit containing both the chain and
the read-out junction. We start by presenting the gate-voltage dependence
of the switching current (Fig. \ref{CentralIslandOscilFig}) for small
variations of the gate voltage (so that $|V_{G}|\ll2e/C_{4}^{g}$)
and for two different ratios of $E_{J}/E_{C}$ . As the central coupling
$C_{3}^{g}$ is $\sim10$ times larger than any other coupling, the
gate voltage only polarizes the middle island. The values of the island
charges result from the combined effect of the gate voltage and off-set
charges. In the particular case of our circuit, the latter vary randomly
within a time scale of $\sim5$ min in average, enabling the measurement
of a single charge configuration during a gate voltage scan that takes
$\sim3$ minutes. Thus, by repeating the same voltage sweep, we measure
different charge configurations. The results presented in Fig. \ref{CentralIslandOscilFig}
were post-selected from a large set of data, of about $3000$ runs
by choosing the largest observed switching current oscillations. The
largest oscillations displayed in Fig. \ref{CentralIslandOscilFig}
correspond to a particularly simple case in which all charges, except
the one on the middle island are close to zero: $\kappa'=(0\pm0.1,0\pm0.1,\frac{C_{3}^{g}V_{g}}{2e},0\pm0.1,0\pm0.1)$.
Notice that the \emph{ab-initio }probability to produce this charge
configuration is $0.2^{4}=0.0016$ which translates into 6 configurations
out of 3000, so its observation supports the assumption of random
charge distribution.

The charge frustration on the middle island introduces a geometrical
phase shift $\exp\left[i2\pi q_{3}\right]$ between the three CQPS
occurring on the junctions at the left side of the middle island and
the three CQPS on the junctions at the right of the middle island.
This phase shift between the different CQPS is graphically represented
for several charge configurations in Fig. \ref{CentralIslandOscilFig}a.
In this regime the total CQPS amplitude (eq. \ref{eq:QPS}) as a function
of gate voltage becomes:\begin{equation}
|v^{*}|=3v\sqrt{2+2\cos\left(\pi V_{g}C_{3}^{g}/e\right)}.\label{Eq: QPS amplitude vs Central Island Vg}\end{equation}

The phase-slip amplitude is expected to vanish completely $v^{*}=0$
for the charge configuration $(0,0,0.5,0,0)$ while the maximum value
$v^{*}=6v$ is obtained for the charge configuration $(0,0,0,0,0)$.
The red line of Fig. \ref{CentralIslandOscilFig} shows the corresponding
theoretical calculation using the CQPS model (\ref{eq:TightBindingHamiltonian}),(\ref{Eq: QPS amplitude vs Central Island Vg}).
Around the charge configuration $(0,0,0.5,0,0)$ we expect a complete
suppression of the total phase-slip amplitude $v^{*}$ (see Fig. \ref{CentralIslandOscilFig}a),
hence an increase of the supercurrent through the chain. For $E_{J}/E_{C}=3$,
the change in the measured switching current due to the full suppression
of CQPS is $\mbox{\ensuremath{\sim}1nA}$, which represents $\sim2\%$
of the critical current of one SQUID, $i_{0}$. Increasing the CQPS-amplitude
by decreasing the ratio $E_{J}/E_{C}$ to a value of $2$ , the oscillation
amplitude of the switching current increases to $\sim2nA$ which represents
$\sim5\%$ of $i_{0}$ (see Fig. \ref{CentralIslandOscilFig}c).

\begin{figure}[tbph]
\begin{centering}
\includegraphics[width=0.9\columnwidth]{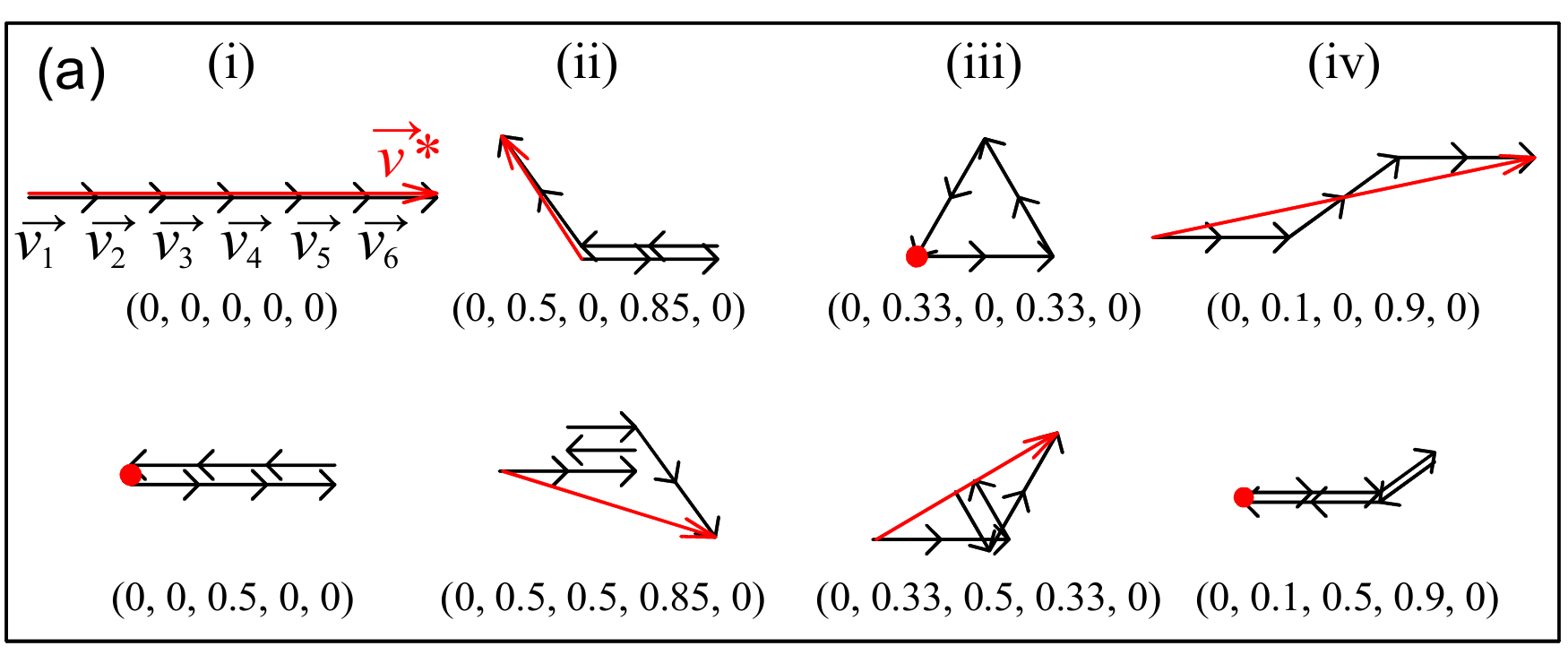}
\par\end{centering}

\begin{centering}
\includegraphics[width=0.9\columnwidth]{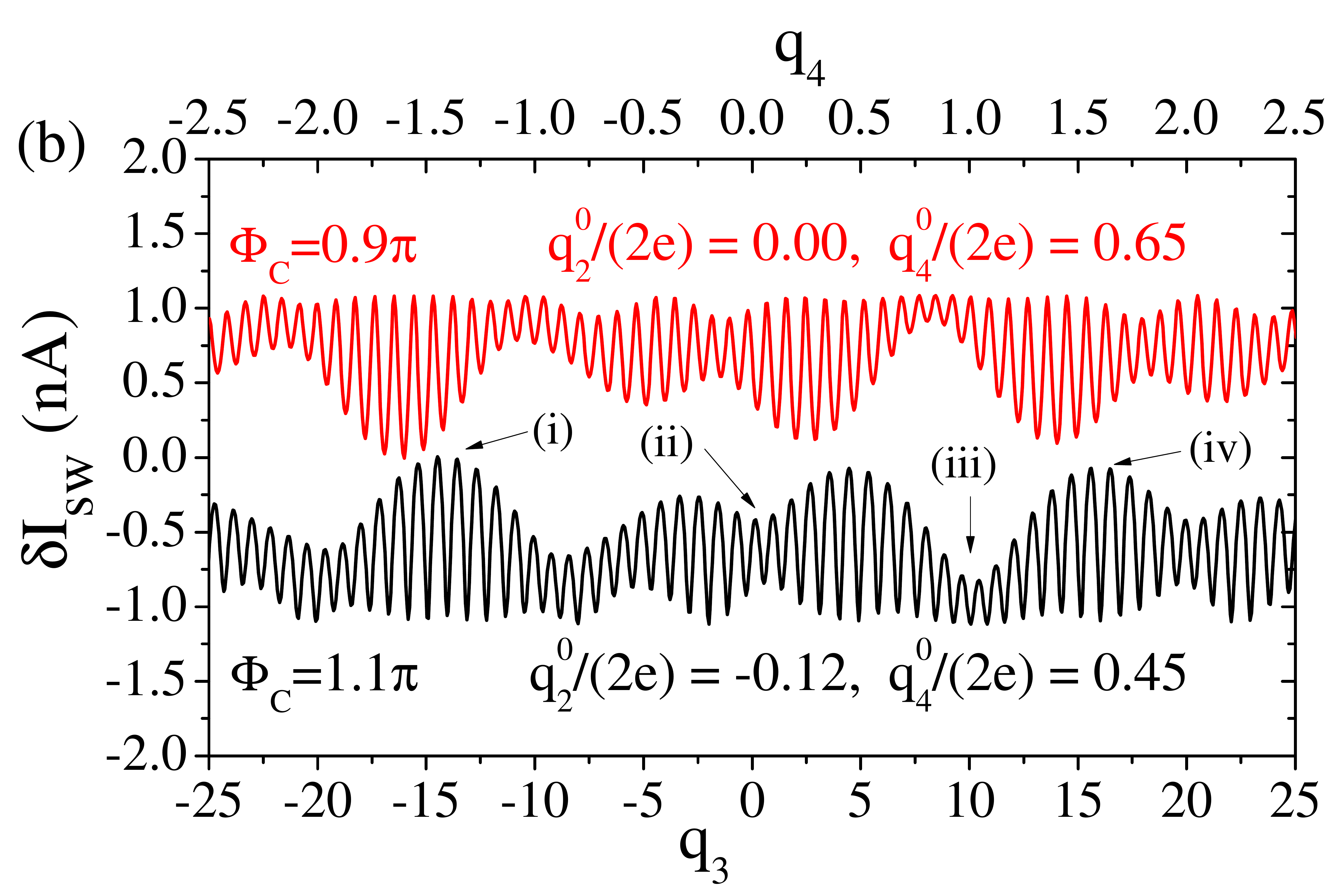}
\par\end{centering}

\begin{centering}
\includegraphics[width=0.9\columnwidth]{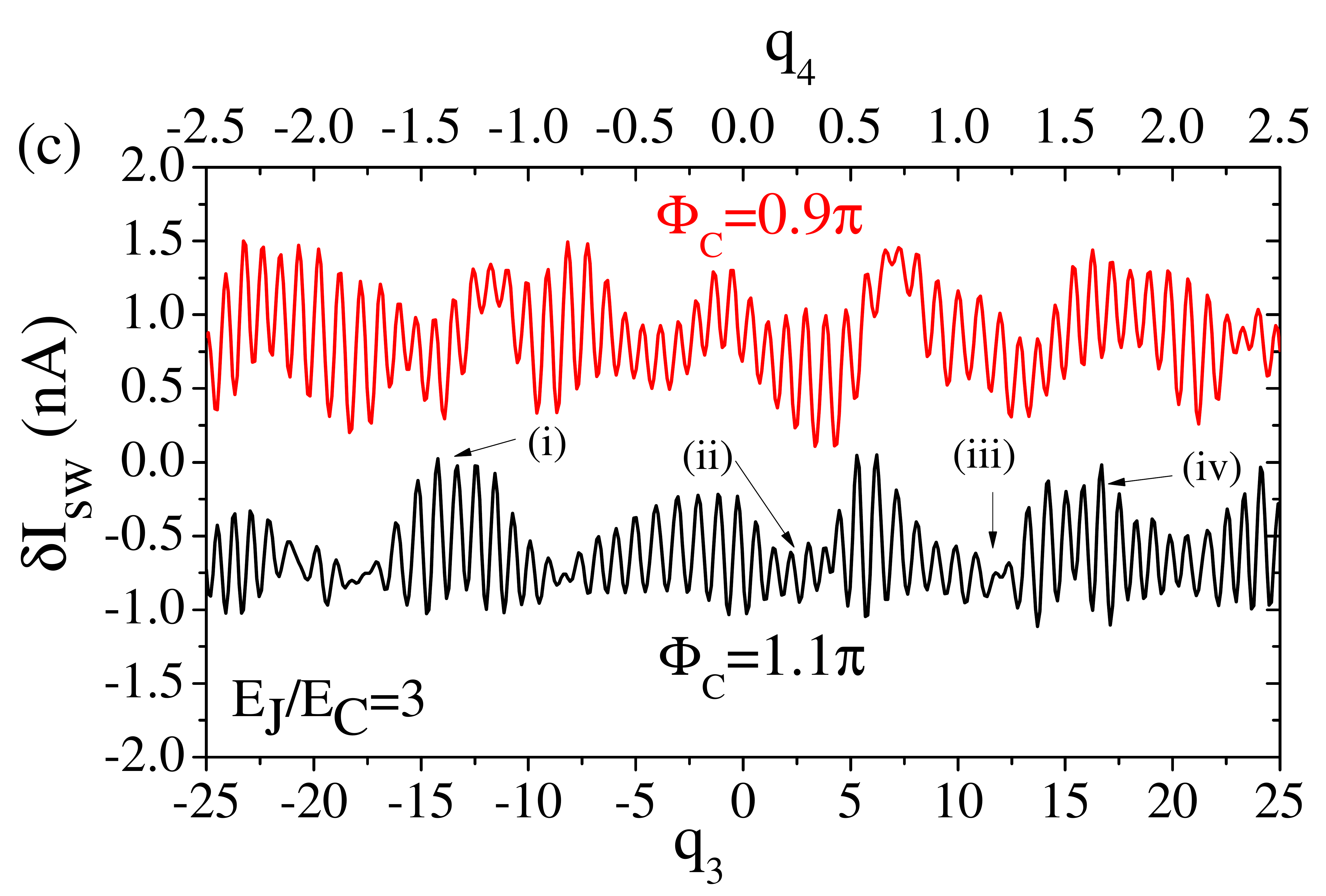}
\par\end{centering}

\caption{CQPS interferences induced by the polarization charge on the middle
and the first two lateral islands $\left(0,q_{2},q_{3},q_{4},0\right)$
of the 6 Josephson junction chain at $E_{J}/E_{C}=3$. (a) Schematic
representations of the 6 phase-slip probability amplitudes $v_{i}$
(in black) and the total CQPS amplitude $v^{*}$ (in red) as vectors
in the complex plane, for several particularly chosen charge configurations.
(b) The calculated switching current oscillation, $\delta I_{SW}$,
induced by the polarization charges for a large sweep of the gate
voltage $V_{G}$, at two different phase biases. The polarization
charge on the central island $q_{3}$ is shown on the lower \textit{x-axis}
and the charge on one of the lateral islands $q_{4}$ is shown on
the higher \textit{x-axis.} (c) The measured $\delta I_{SW}$ over
a large sweep of $V_{G}$ at the same phase-biases $\Phi_{C}$ as
in (b). The value of $\Phi_{C}$ for each curve is shown on the right
side of the figure. }

\label{Fig: CompleteInterferencePatern}
\end{figure}

We now turn to the discussion of more complex interferences of CQPS
realized by increasing the gate voltage ($|V_{G}|\gtrsim2e/C_{4}^{g}$)
that leads to the polarization of the islands next to the central
island. In Fig. \ref{Fig: CompleteInterferencePatern}c we show two
interference patterns that were post-selected from a total of 200
curves. Again, the selection criteria was the maximum observed switching
current amplitude. In the following we show that these measured patterns
can be understood by considering charge configurations in which only
$q_{1},q_{5}\approx0$: \foreignlanguage{french}{$\kappa''=(0\pm0.1,q_{2}^{(0)}+\frac{C_{2}^{g}V_{g}}{2e},q_{3}^{(0)}+\frac{C_{3}^{g}V_{g}}{2e},q_{4}^{(0)}+\frac{C_{4}^{g}V_{g}}{2e},0\pm0.1)$}.
The corresponding \textit{ab-initio} probability is $0.2^{2}=4\%$
which translates into $\sim8$ curves out of the measured $200$.

The charge frustration on the middle, the second and the fourth islands
introduces geometrical phase shifts between the CQPS of the second,
third and forth junctions. Fig. \ref{Fig: CompleteInterferencePatern}a
show the corresponding CQPS amplitudes as vectors in the complex plane
for several charge configurations. The resulting switching current
oscillations $\delta I_{SW}$, presented in Fig. \ref{Fig: CompleteInterferencePatern}b,
show a complex pattern, composed of a fast harmonic arising from the
strong $C_{3}^{g}$ coupling and a slower evolving envelope due to
the weaker $C_{2}^{g}$ and $C_{4}^{g}$ couplings (see Table \ref{Flo:chFitParam}).
In Fig. \ref{Fig: CompleteInterferencePatern}c we show the measured
interference patterns for two different phase biases $\Phi_{C}$ of
the Josephson junction chain; these biases were chosen close to $\Phi_{C}=\pi$
in order to maximize the response of the chain. For the top curves
in Fig. \ref{Fig: CompleteInterferencePatern}b and c we polarized
the chain at $\Phi_{C}\lesssim\pi$ so we expect the switching current
to increase when the phase slips are suppressed. Similarly, for the
bottom curves, where $\Phi_{C}\gtrsim\pi$, we expect the switching
current to decrease when the chain becomes classical. The exact shape
of the oscillations envelope depends on the configuration of the offset
charges $q_{i}^{0}$. For the two calculated curves we have chosen
the offset charges configurations $q_{2}^{(0)},q_{4}^{(0)}$ that
give the best fit the experimental data. The exact values of the fit
parameters are shown in table. \ref{Flo:chFitParam}.

\begin{table}[h]
\begin{centering}
\begin{tabular}{|c|c|c|c|c|c|}
\hline
\multicolumn{1}{|c|}{$\Phi_{C}$} & $q_{2}^{0}$$(2e)$ & $q_{4}^{0}$$(2e)$ & $C_{2}^{g}$$(aF)$ & $C_{3}^{g}$$(aF)$ & $C_{4}^{g}$$(aF)$\tabularnewline
\hline
\hline
$0.9\pi$ & $0$ & $0.65$ & $25$ & $410$ & $42$\tabularnewline
\cline{1-3}
$1.1\pi$ & $0.12$ & $0.45$ &  &  & \tabularnewline
\hline
\end{tabular}
\par\end{centering}

\caption{Fit parameters for the calculated QPS interference patterns presented
in Fig. \ref{Fig: CompleteInterferencePatern}b}

\label{Flo:chFitParam}
\end{table}

The qualitative behavior of the interference pattern agrees perfectly
with the theoretical expectations based on the simple picture of addition
of complex amplitudes. Using the fitted values of $q_{i}^{(0)}$ we
evaluate the gate voltages corresponding to the special charge configurations
shown in Fig. \ref{Fig: CompleteInterferencePatern}a and indicated
them on the measured and calculated curves. For instance, point $(i)$
corresponds to the configuration where all charges are zeros, the
same configuration that was realized in measurements displayed in
Fig. \ref{CentralIslandOscilFig}a at $q_{3}=0$. Notice that in the
measurements presented in Fig. \ref{Fig: CompleteInterferencePatern}
its position is shifted along the \textit{x axis} by the offset charges
on the islands $2$ and $4$. Because for arbitrary values of $q_{i}$
the total CQPS amplitude is less than $Nv$, one expects that maximal
oscillations as a function of $q_{3}$ occur around point $(i)$ (see
Fig. \ref{Fig: CompleteInterferencePatern}a). Indeed, next to this
point, the switching current $\delta I_{SW}$ oscillations have the
largest amplitude. The chain goes from the perfectly coherent phase-slips
regime at $q_{3}=0$, where the switching current is minimum (the
zero level in Fig. \ref{CentralIslandOscilFig}b and c), to the maximally
dephased configuration at $q_{3}=0.5$ when the phase slips are canceled,
the chain is almost classical and the critical current is enhanced.
Point $(ii)$ corresponds to opposite limit in which the oscillations
are strongly suppressed due to interference induced by non-zero charges
at $q_{2},q_{3}.$ Working point $(iii)$ shows the situation when
the total CQPS amplitude is suppressed at $q_{3}=0$ and it never
reaches the maximum $Nv$ for any value of $q_{3}$ (see Fig. \ref{Fig: CompleteInterferencePatern}a).
In this case we expect small oscillations of $\delta I_{SW}$ that
reach the maximum supercurrent for the classical chain. The case of
$(iv)$ shows that it is not necessary to have the CQPS perfectly
aligned as in $(i)$ in order to have a large amplitude of $\delta I_{SW}$
oscillations. As expected, the $\delta I_{SW}$ oscillations around
$(iv)$ are comparable in amplitude to the ones around point $(i)$.

The Aharonov-Casher interference of phase slips is expected to be
a $2e$ periodic effect. In general, random $1e$ quasi-particle poisoning
is a severe problem for the observation of the interference effect
as the $1e$ contamination reduces the accessible charge interval
from $\left[0,2e\right]$ to $\left[0,1e\right]$. As a consequence,
the amplitude of the oscillations in the interference pattern is significantly
reduced by $1e$ quasi-particle poisoning\cite{Elion1993}. In our
case we observe the quasi-particle poisoning by the appearance of
a $1e$ periodicity in the $\delta I_{SW}$ \textit{vs. $V_{G}$}
oscillations for temperatures above $T=300\, mK$. At base temperature
$T=50\, mK$ we scan the full charge space interval $\left[0,2e\right]$
enabling the observation of complete destructive CQPS interference
(see Fig. \ref{CentralIslandOscilFig}). Furthermore, spectroscopy
measurements of propagative modes in long SQUID chains \cite{PopThesis2011}
independently confirms the value of the gate capacitance $C_{g}^{3}\simeq400pF$,
corresponding to a $2e$ periodic $\delta I_{SW}$ \textit{vs. $V_{G}$
}curve at low temperature.

As can be seen from the periodic dependence of the measured switching
current, the island charge configuration does not change significantly
during the measurement. Although each measurement point implies $10^{4}$
repeated switchings into the dissipative state of the junctions, where
large numbers of quasi-particles are excited, after the circuit relaxes
back to the dissipationless state, the charge configuration is stable
enough in order to enable the measurement of the interference pattern.
We have directly measured the frequency of random charge jumps by
repeating the same measurement several times and we observe a typical
time of $\tau_{qp}\sim5$ minutes between changes in the island charge
configuration. This time interval is sufficient in our case as it
enables the measurement of several hundreds of experimental data points.

It is well established that slow drift of charge induced by fluctuating
TLS leads to $\delta q\ll e$ at time scales of minutes\cite{Zimmerman2008}.
A significant charge drift at this and smaller time scales is attributed
to non-equilibrium quasi-particles jumping between the islands\cite{Martinis2009};
the equilibration of these quasi-particles is made difficult by their
localization in subgap states\cite{Faoro2008}. Surprisingly, the
time scales of these jumps might be dramatically different even in
similar devices. Charge fluctuation times, similar to the one observed
here, have been reported previously in small highly resistive Josephson
islands\cite{Maisi2009}, in small charge-phase qubits\cite{Vion2002,Fay2008}
and in Cooper-pair transistors\cite{Lafarge1993,Bibow2002}. However
much shorter times, $\tau_{qp}<1s$, were reported for the fluxonium
circuit\cite{Manucharyan2011} and even shorter times, $\tau_{qp}\sim1\mu s$
, were reported for larger devices such as the transmon\cite{Shreier2008}.

Reasonably good agreement between our measurements and a model of
independent phase slips might be surprising given the modest values
of $E_{J}/E_{C}=2-3$ of our Josephson chains, because this approximation
is expected to be correct only for $E_{J}/E_{C}\gg1$\cite{Matveev2002}.
Josephson junction circuits with $E_{J}/E_{C}\sim1$ are typically
analyzed using the charge basis description and only a couple of charging
states are needed for accurate results. At larger $E_{J}/E_{C}$ the
number of required states grows, which makes the problem of numerical
diagonalization difficult, especially for large systems. From a practical
point of view it is important to compare both approaches and their
validity as a function of $E_{J}/E_{C}$, because the calculation
is many orders of magnitude faster in the phase slip approximation,
in particular for systems containing a large number of junctions\cite{Guichard2010}.

We now discuss the details and the validity of the diagonalization
of the full Hamiltonian of the chain:

\begin{multline}
\begin{array}{c}
H=\frac{4e^{2}}{2}\sum_{i,j}\left[C^{-1}\right]_{ij}\left(Q_{i}-q_{i}\right)\left(Q_{j}-q_{j}\right)+\\
+\sum_{i=1}^{6}E_{J}\left[1-\cos\left(\varphi_{i}-\varphi_{i-1}\right)\right]\end{array}\label{eq:TotalHamiltonianJosephsonChain}\end{multline}
where $Q_{i}$ is the charge (in units of $2e$) on the \textit{i-th}
island and $\varphi_{i}$ is the superconducting phase on the island.
$C^{-1}$ is the matrix of inverse capacitance of the chain. The first
sum in the expression (\ref{eq:TotalHamiltonianJosephsonChain}) is
the charging energy for the islands of the chain and the second sum
represents the Josephson couplings for all junctions in the chain.
As the total phase difference $\gamma$ across the chain is fixed,
we have $\varphi_{0}=0$ and $\varphi_{6}=\gamma$. The charges $Q_{i}$
are multiples of the elementary charge of a Cooper pair. As $Q_{i}$
and $\varphi_{i}$ are conjugate variables, the chain's wavefunction
$\psi(\varphi_{1},...,\varphi_{5})$ is $2\pi$ periodic in $\varphi_{i}$.
From the Hamiltonian (\ref{eq:TotalHamiltonianJosephsonChain}) one
can see that its energy spectrum is a periodic function of the polarization
charges $q_{i}$: indeed, any modification of the polarization charge
by the charge of a Cooper pair $q_{i}\rightarrow q_{i}+1$ can be
absorbed by the unitary transformation $\exp(i\hat{\varphi}_{i})$
which changes $Q_{i}$ into $Q_{i}+1$, while leaving $\varphi_{i}$
invariant. Therefore the supercurrent through the chain $I\left(\varphi_{1},...,\varphi_{6}\right)$
remains unchanged when the polarization charges change by a multiple
of the charge of a Cooper pair.

The blue curves in Fig. \ref{CentralIslandOscilFig}b and c show the
calculated switching current oscillations, from the numerical diagonalization
of the total Hamiltonian (\ref{eq:TotalHamiltonianJosephsonChain})
(for details see the Methods section), as a function of $V_{G}$ for
the charge configurations $\kappa'$. These calculations agree reasonably
well with the switching current calculated using the semi-classical
CQPS approximation (the MLG model). The modulation of the critical
current expected theoretically is somewhat larger than the data. This
is due to the fact that in the experiment the random charges $q_{i}$
are not exactly zeros; this results in the interference that decreases
the observed amplitude of the switching current modulation as we discussed
above for the charge configurations $\kappa''$.

In conclusion, we have presented a quantitative study of the Aharonov-Casher
effect exhibited by fluxon motion in a multi-junction circuit. We
compared the data with the expectations based on the diagonalization
of the full Hamiltonian of the chain in the charge basis. Our results
show that the ground state properties of a short Josephson junction
chain can be fully understood in terms of phase slip dynamics even
in a parameter range that has been traditionally described in terms
of charge dynamics. The measurements also show that the polarization
charges on the islands of the chain can be controlled with sufficient
precision and they are stable enough to enable the observation of
the chain's collective behavior at the time scale of minutes. We believe
that our results will provide a starting point to reconsider the physics
of large Josephson junction arrays, long Josephson chains and their
possible applications to the frequency-to-current conversion device
or a topologically protected qubit.

We would like to thank B. Pannetier, Daniel Esteve, Frank Hekking,
Gianluca Rastelli and Christoph Schenke for fruitful discussions.
The research has been supported by the European STREP MIDAS and the
French ANR QUANTJO. L. Ioffe acknowledges support from NSF ECS-0608842,
ARO 56446-PH-QC and DARPA HR0011-09-1-0009. I. Protopopov acknowledges
support from the Alexander von Humboldt Foundation and the DFG Center
for Functional Nanostructures.

\subsection*{Methods}

\paragraph*{Switching current measurements}

The switching current was determined from the switching probability
at $50$\%. We apply $\sim10^{4}$ bias-current pulses of amplitude
$I_{bias}$ and measure the switching probability as the ratio between
the number of switching events and the total number of pulses. The
measured switching current corresponds to the escape process out of
the total potential energy containing the contributions of the read-out
junction and the chain. We can calculate this escape process and therefore
deduce the effect of quantum phase-slips on the ground state of the
chain \cite{Pop2010}. From the escape rate, we can deduce the escape
probability $P_{SW}$ as a function of the current bias $I_{bias}$
and infer the theoretical switching current (at $P_{SW}=50\%$) for
each biasing point $\left(\Phi_{C},V_{G}\right)$.

\paragraph*{Numerical diagonalization of the exact chain Hamiltonian}

The Hamiltonian (\ref{eq:TotalHamiltonianJosephsonChain}) gives the
exact description of the Josephson circuit (provided that all energy
scales remain small compared to Cooper gap $\Delta$ and no other
degrees of freedom are involved. For the purposes of numerical diagonalization
one has to limit the number of charging states on each island. This
approximation can be easily tested by comparing the results of diagonalization
for different number of allowed charging states. For the problem here
with $E_{J}/E_{C}=2-3$ it is sufficient to keep 7 charging states
to get the results with $10^{-2}$ accuracy.

\subsection*{Supplementary information: The hopping term in the Matveev-Larkin-Glazman
theory of quantum fluctuations}

Here we present the detailed derivation of the hopping term $v^{*}$
of the MLG model in the charge frustrated chain. Similar calculations
have been performed for the Josephson chain \cite{Matveev2002} and
for slightly different Josephson circuits \cite{Friedman2002,Ivanov2002}.
To calculate the hopping term we need to find the classical trajectories
connecting states before and after one phase-slip event. There are
$N$ such trajectories, each of them corresponding to the phase slip
occurring on a particular junction in the chain. In a semi classical
approximation, the contribution of the phase slip in the junction
$i$ to the hopping term is governed by the imaginary-time action
$S_{i}$ on the corresponding trajectory: \begin{equation}
v_{i}=Ae^{-S_{i}}\label{eq:HoppingTermOn Junction_i}\end{equation}
The prefactor $A$ accounts for the contribution of the non-classical
trajectories close to the classical one that defines $S_{i}$.

In order to calculate the actions $S_{i}$, we need to derive the
complete Lagrangian for the Josephson chain. The electrostatic effects
in the Josephson chain are described by the following Hamiltonian:

\begin{equation}
H_{C}=\frac{1}{2}\sum_{i,j}\left[C^{-1}\right]_{ij}\left(Q_{i}-q_{i}\right)\left(Q_{j}-q_{j}\right)\label{ElectrostaticHamilt}\end{equation}

The polarization charges $q_{i}=\frac{C_{i}^{g}V_{g}}{2e}$ are controlled
by the gate voltage. We would like to mention that in our experimental
setup we have added screening lines to the central gate, in order
to obtain a coupling to the central island at least 10 times larger
than the couplings to the rest of the chain: $C_{3}^{g}\simeq10*C_{4}^{g},C_{2}^{g}\simeq50*C_{1}^{g},C_{5}^{g}$.

Since the charges $Q_{i}$ and the phases of the islands $\varphi_{i}$
are canonical conjugate variables, the equation of motion for the
phase reads:

\begin{equation}
\dot{\varphi}_{i}=\frac{\partial H_{C}}{\partial Q_{i}}=\sum_{j}\left[C^{-1}\right]_{ij}\left(Q_{j}-q_{j}\right)\quad\Longrightarrow\quad Q_{i}=\sum_{j}C_{ij}\dot{\varphi_{j}}+q_{i}\label{eq:EqOfMotionForPhase}\end{equation}

Using eq. (\ref{eq:EqOfMotionForPhase}) we can rewrite the charging
Hamiltonian (\ref{ElectrostaticHamilt}) in the phase notation:

\begin{equation}
H_{C}=\frac{1}{2}\sum_{i,j}C_{ij}\dot{\varphi_{i}}\dot{\varphi_{j}}\label{eq:Charge Hamiltonian in phase basis}\end{equation}

The charge part of the Lagrangian for the Josephson junction chain
reads:

\begin{equation}
\mathcal{L_{C}}=\sum_{i}Q_{i}\dot{\varphi_{i}}-H_{C}\label{eq:ChargeLagrangianByDef}\end{equation}

Following formula (\ref{eq:ChargeLagrangianByDef}) and using the
expressions (\ref{eq:EqOfMotionForPhase}) and (\ref{eq:Charge Hamiltonian in phase basis})
we get for the charge Lagrangian the following expression:\[
\mathcal{L_{C}}=\sum_{ij}C_{ij}\dot{\varphi_{i}}\dot{\varphi_{j}}+\sum_{i}q_{i}\dot{\varphi_{i}}-\frac{1}{2}\sum_{ij}C_{ij}\dot{\varphi_{i}}\dot{\varphi_{j}}\]

\begin{equation}
\mathcal{L_{C}}=\frac{1}{2}\sum_{ij}C_{ij}\dot{\varphi_{i}}\dot{\varphi_{j}}+\sum_{i}q_{i}\dot{\varphi_{i}}\label{eq:ChargeLagrangianInPhaseNotation}\end{equation}

The capacitance matrix $C_{ij}$ contains the values of all coupling
between the islands. However, in reality, due to the geometry of the
sample, the capacitance between first neighbors is orders of magnitude
larger then the stray capacitance between second order neighbors.
This means that we can safely work within the so called \textit{nearest
neighbor capacitance} approximation, and the matrix $C_{ij}$ only
gets non zero contributions for the elements closest to the main diagonal:
\begin{equation}
\left(\begin{array}{ccccc}
2C & -C & 0 & ... & 0\\
-C & 2C & -C & ... & 0\\
0 & -C & 2C & ... & 0\\
... & ... & ... & ... & -C\\
0 & 0 & 0 & -C & 2C\end{array}\right)\label{eq:Capacitance matrix}\end{equation}
Where $C$ is the capacitance of one junction in the chain.

Using the approximation (\ref{eq:Capacitance matrix}) the expression
of the charge Lagrangian is simplified and it reads:

\begin{equation}
\mathcal{L_{C}}=\frac{1}{2}\sum_{i}C(\dot{\varphi_{i}}-\dot{\varphi_{i-1}})^{2}+\sum_{i}q_{i}\dot{\varphi_{i}}\label{eq:ChargeLagrangianNearestNeighbourCapacitanceLimit}\end{equation}

Introducing the phase differences on the junctions $\theta_{i}=\varphi_{i+1}-\varphi_{i}$
and including the Josephson energy, we derive the complete Lagrangian
of the chain:

\begin{equation}
\mathcal{L}=\sum_{i}\left[\frac{\dot{(\theta_{i})}^{2}}{16E_{C}}-E_{J}\cos\theta_{i}\right]-\sum_{i}p_{i}\dot{\theta_{i}}\;,\qquad p_{i}=\sum_{j=1}^{i-1}q_{i}\label{eq:CompleteLagrangianOfJosephsonChain}\end{equation}

We can see that the Lagrangian (\ref{eq:CompleteLagrangianOfJosephsonChain})
has two components which have very different physical consequences.
The first sum that we call $\mathcal{L}_{0}$ is independent on the
frustration charges $q_{i}$. It gives a contribution to the real
part of the phase-slip amplitude $v_{i}$, that is given by the Bloch
band width $v=16\sqrt{\frac{E_{J}E_{C}}{\pi}}\left(\frac{E_{J}}{2E_{C}}\right)^{0.25}e^{-\sqrt{8\frac{E_{J}}{E_{C}}}}$.
For identical junctions in the chain, $v$ is independent on the path
chosen by the phase slip. The second sum of the Lagrangian (\ref{eq:CompleteLagrangianOfJosephsonChain}),
which we call $\mathcal{\delta L}$, has the form of a total time
derivative. Hence, this term does not change the classical equations
of motion and the real part of the classical action on a single trajectory.
However, $\mathcal{\delta L}$ gives the tunneling amplitude along
each path its own phase factor. This phenomenon is mathematically
equivalent to the AB effect for the \textit{phase variable} $\varphi_{i}$
which is $2\pi$ periodic. Changing $p_{i}$ amounts to changing the
periodic boundary conditions $\psi(\varphi_{i}+2\pi)=e^{i2\pi p_{i}}\psi(\varphi_{i})$
for the phase, in analogy to the motion of a charged particle on a
circle, threaded by a flux tube.

When a phase-slip occurs on junction $i$, the other phase differences
$\theta_{j}$ are changed by:

\begin{equation}
\Delta\theta_{j}=-\frac{2\pi}{N}+2\pi\delta_{ij}\label{eq:PhaseSlipJumps}\end{equation}
Thus, the contribution to the phase-slip action from the\textit{ j-th}
junction in the presence of charge frustration reads:

\begin{equation}
\delta S_{j}=-i\int\mathcal{\delta L\,}dt=-i\sum_{k}p_{k}\Delta\theta_{k}=-2\pi ip_{j}-\frac{2\pi i}{N}\sum_{k}p_{k}\label{eq:ActionDueToChargeFrustration}\end{equation}
Since the last term in the expression above does not depend on \textit{$k$,}
it only adds an overall phase term for all phase slip trajectories,
thus has no physical effect on the interference pattern and it can
be dropped. Replacing this result in the formula (\ref{eq:HoppingTermOn Junction_i}),
we get the mathematical expression for the charge frustration dephasing
factor in the phase slip probability amplitude of the \textit{j-th}
junction:

\begin{equation}
\delta v_{j}=e^{i2\pi p_{j}}\label{eq:ChargeFrustrationDephasingFactor}\end{equation}

So the phase slip probability amplitude on the\textit{ j-th} junction
$v_{j}$ reads:

\begin{equation}
v_{j}=v\, exp\left[i2\pi\sum_{k=1}^{j-1}q_{k}\right]\label{QPSprobAmp}\end{equation}

In other words, the absolute value of the probability amplitude for
the QPS is the same as in the absence of charge frustration, but the
geometric phase difference between the QPS is proportional to the
total charge on the islands between the junctions. Finally, the full
hopping term between the states $|m\rangle$ and $|m+1\rangle$ in
the presence of charge frustration is the sum of phase slip amplitudes
$v_{i}$ in all six junctions:

\begin{equation}
v^{*}=\sum_{i=1}^{6}v_{i}\label{TotalCouplingTerm}\end{equation}

At zero gate voltage, the expression (\ref{TotalCouplingTerm}) reduces
to $v^{*}=Nv$ that was used in the previous section to solve the
tight binding MLG Hamiltonian and calculate the expected switching
current. Non-zero gate voltage directly affects the interference of
QPS by changing the geometrical Aharonov-Casher phase difference between
phase slips in different junctions and thus provides a direct test
for the quantum nature of the chain's ground state.

\end{document}